\documentclass[prl,twocolumn,twoside,showpacs,reprint]{revtex4-1}
\pdfoutput=1
\usepackage{graphicx,amssymb,amsmath,epsf,bm}
\epsfclipon
\newcommand{\p}{\partial}

\newcommand{\unit}[1]{~\mathrm{#1}}
\renewcommand{\(}{\left(}
\renewcommand{\)}{\right)}
\renewcommand{\[}{\left[}
\renewcommand{\]}{\right]}

\newcommand{\expct}[1]{\langle{#1}\rangle}
\newcommand{\Expct}[1]{\left\langle{#1}\right\rangle}
\newcommand{\cum}[1]{\langle{#1}\rangle_\mathrm{c}}

\newcommand{\prt}[2]{\frac{\partial{#1}}{\partial{#2}}}
\newcommand{\prts}[3]{\frac{\partial^{#3}#1}{\partial{#2}^{#3}}}

\renewcommand{\eqref}[1]{Eq.~(\ref{#1})}
\newcommand{\eqsref}[1]{Eqs.~(\ref{#1})}
\newcommand{\pref}[1]{(\ref{#1})}
\newcommand{\figref}[1]{Fig.~\ref{#1}}
\begin{document}

\title{Crossover from Growing to Stationary Interfaces in the Kardar-Parisi-Zhang Class}

\author{Kazumasa A. Takeuchi}
\email{kat@kaztake.org}
\affiliation{Department of Physics,\! The University of Tokyo,\! 7-3-1 Hongo,\! Bunkyo-ku,\! Tokyo 113-0033,\! Japan}%

\date{\today}

\begin{abstract}
This Letter reports on how the interfaces
 in the (1+1)-dimensional Kardar-Parisi-Zhang (KPZ) class undergo,
 in the course of time, a transition from the flat,
 growing regime to the stationary one.
Simulations of the polynuclear growth model
 and experiments on turbulent liquid crystal
 reveal universal functions of the KPZ class
 governing this transition,
 which connect
 the distribution and correlation functions
 for the growing and stationary regimes.
This in particular shows
 how interfaces realized in experiments and simulations
 actually approach the stationary regime, 
 which is never attained unless a stationary interface
 is artificially given as an initial condition.
\end{abstract}


\maketitle

Aside from their ubiquity in nature,
 surface growth phenomena constitute an important situation
 of statistical mechanics out of equilibrium,
 where scale invariance and universal scaling laws
 arise generically
 \cite{Barabasi.Stanley-Book1995}.
These are usually evidenced in the roughness of the interfaces,
 whose amplitude $w(L,t)$ measured at the system (substrate) size $L$
 and time $t$ obeys the following power laws:
\begin{equation}
 w(L,t) \sim \begin{cases} L^\alpha & \text{for $L\ll{}L_*$,} \\ t^\beta & \text{for $L\gg{}L_*$,} \end{cases} \quad (L_* \sim t^{1/z}),  \label{eq:FV}
\end{equation}
 with scaling exponents $\alpha,\beta,z\equiv\alpha/\beta$
 \cite{Family.Vicsek-JPA1985,Barabasi.Stanley-Book1995}.
At the heart of such growth processes is
 the Kardar-Parisi-Zhang (KPZ) equation \cite{Kardar.etal-PRL1986}
 and the corresponding universality class
 \cite{Barabasi.Stanley-Book1995,Kardar.etal-PRL1986},
 describing the simplest case without any conservation laws
 and long-range interactions.
For one-dimensional interfaces,
 the KPZ equation reads
\begin{equation}
 \prt{}{t}h(x,t) = \nu \prts{h}{x}{2} + \frac{\lambda}{2}\(\prt{h}{x}\)^2 + \sqrt{D} \eta(x,t),  \label{eq:KPZEq}
\end{equation}
 where $h(x,t)$ denotes the fluctuating height profile
 and $\eta(x,t)$ white Gaussian noise with $\expct{\eta(x,t)}=0$ and
 $\expct{\eta(x,t)\eta(x',t')}=\delta(x-x')\delta(t-t')$.
The values of the scaling exponents are exactly known
 in this one-dimensional case
 \cite{Barabasi.Stanley-Book1995,Kardar.etal-PRL1986,Forster.etal-PRA1977}:
 the height fluctuation $\delta{}h\equiv{}h-\expct{h}$
 grows as $\delta{}h\sim{}t^{1/3}$ ($\beta=1/3$)
 and the correlation length $\xi$ as $\xi\sim{}t^{2/3}$ ($z=3/2$).
Specifically,
 $h$ is described by a rescaled random variable $\chi(x',t)$ as
\begin{equation}
 h(x,t) \simeq v_\infty t + (\Gamma t)^{1/3} \chi(x',t)  \label{eq:Height}
\end{equation}
 with a rescaled coordinate $x'\equiv(Ax/2)(\Gamma{}t)^{-2/3}$
 and constant parameters
 $A\equiv\nu/2D,\Gamma\equiv{}A^2\lambda/2$, and $v_\infty$.
The KPZ-class exponents have indeed been reported
 in various models and theoretical situations
 \cite{Barabasi.Stanley-Book1995,Kardar.etal-PRL1986,Forster.etal-PRA1977, [{For recent reviews, see, e.g., }] Kriecherbauer.Krug-JPA2010, *Sasamoto.Spohn-JSM2010, *Corwin-RMTA2012}
 as well as by a growing number of experiments
 \cite{Wakita.etal-JPSJ1997,Maunuksela.etal-PRL1997,*Myllys.etal-PRE2001,Takeuchi.Sano-PRL2010,*Takeuchi.etal-SR2011,Takeuchi.Sano-JSP2012,Huergo.etal-PRE2010,*Huergo.etal-PRE2011,Yunker.etal-PRL2013,Aegerter.etal-PRE2003}.

Studies on the (1+1)-dimensional KPZ class entered
 an unprecedented stage in 2000,
 when
 Johansson \cite{Johansson-CMP2000}
 and others \cite{Kriecherbauer.Krug-JPA2010}
 rigorously derived asymptotic distributions of the height fluctuations
 for a few models.
Among others, it has brought about two outstanding outcomes.
(i) The KPZ class splits into a few subclasses
 according to the global geometry of the interfaces, or, equivalently,
 to the initial condition.
These subclasses are characterized
 by different distribution and correlation functions,
 whereas they share the same scaling exponents.
(ii) An unexpected link to random matrix theory has been revealed.
In particular, the asymptotic distribution of $\chi$
 for the flat and curved interfaces is given
 by the largest-eigenvalue distribution,
 called the Tracy-Widom (TW) distribution
 \cite{Tracy.Widom-CMP1994,*Tracy.Widom-CMP1996,Mehta-Book2004},
 for the Gaussian orthogonal ensemble (GOE)
 and the Gaussian unitary ensemble,
 respectively \cite{Prahofer.Spohn-PRL2000}.
The stationary interfaces also form a distinct subclass.
To study it analytically,
 one usually sets the initial condition $h(x,0)$ to be a stationary interface,
 which is simply the one-dimensional Brownian motion
 for the KPZ equation \cite{Barabasi.Stanley-Book1995}.
The height difference $h(x,t)-h(x,0)$ then grows as \eqref{eq:Height} and
 $\chi$ obeys the $F_0$ distribution introduced by Baik and Rains
 \cite{Baik.Rains-JSP2000},
 as proved for the polynuclear growth (PNG) model
 \cite{Baik.Rains-JSP2000,Prahofer.Spohn-PRL2000},
 for the totally asymmetric simple exclusion process
 \cite{Ferrari.Spohn-CMP2006,Baik.etal-a2012},
 and, very recently, for the KPZ equation
 \cite{Imamura.Sasamoto-PRL2012,*Imamura.Sasamoto-JSP2013}.
The two-point correlation function
 being exactly derived as well
 \cite{Prahofer.Spohn-JSP2004,Ferrari.Spohn-CMP2006,Baik.etal-CPAM2010,Baik.etal-a2012,Imamura.Sasamoto-PRL2012},
 this subclass
 is now firmly established
 like the ones for the flat and curved interfaces.

\begin{figure*}[t!]
 \includegraphics[width=\hsize,clip]{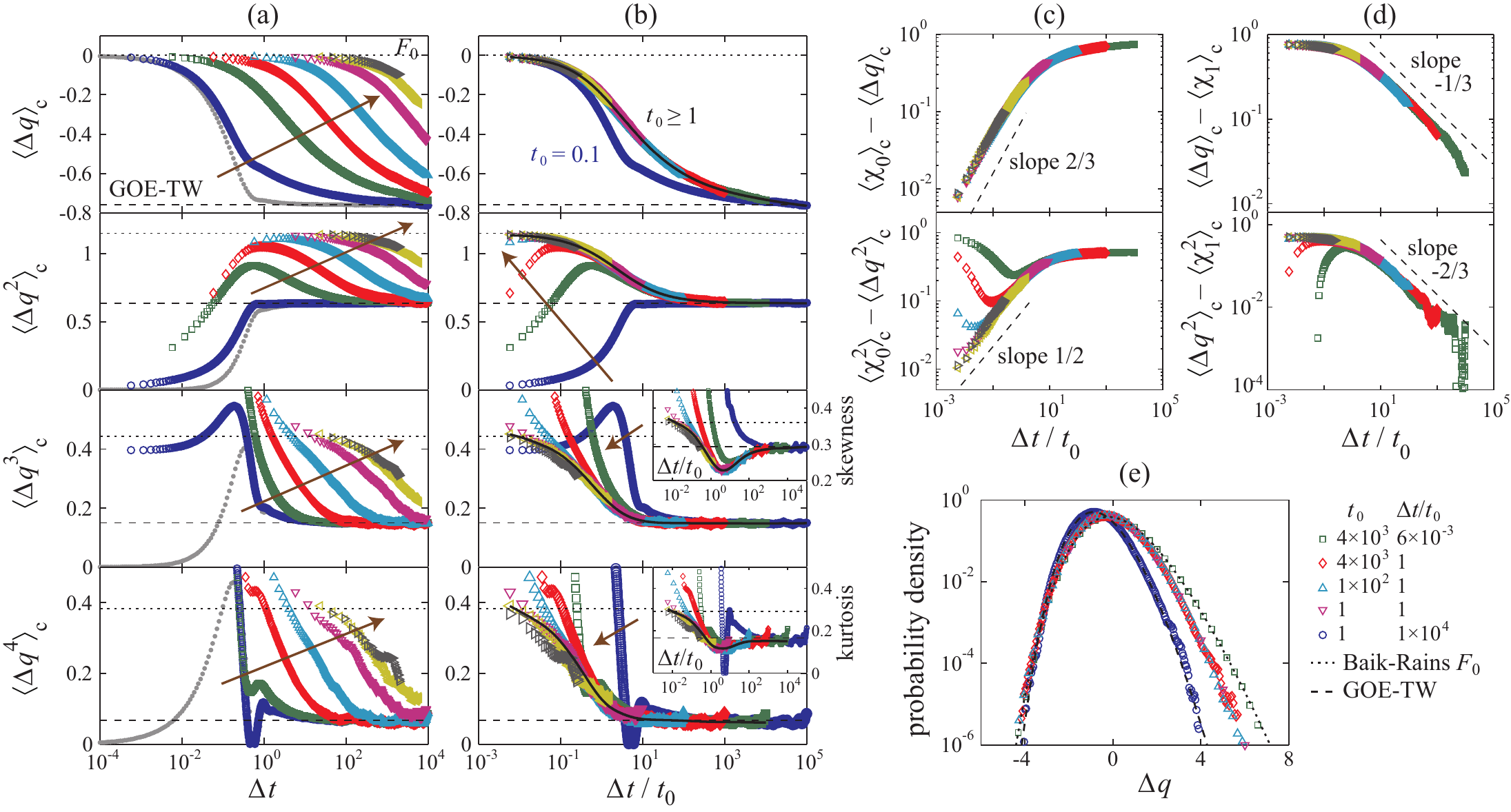}
 \caption{
(color online). 
Crossover in the one-point distribution for the PNG model. 
(a,b) First- to fourth-order cumulants $\cum{\Delta{}q^n}$
 against $\Delta{}t$ (a) and $\Delta{}t/t_0$ (b),
 for $t_0=0,0.1,1,10,100,1000,4000,7000$
 (increasing as the arrows indicate). 
The top (dotted) and bottom (dashed) horizontal lines
 indicate the values
 for the Baik-Rains $F_0$ and GOE-TW distributions,
 $\cum{\chi_0^n}$ and $\cum{\chi_1^n}$, respectively. 
The black solid lines in (b) show fitting to the collapsed curves (see text).
The insets in (b) show the skewness and the kurtosis. 
(c,d) Asymptotic behavior of the data in (b)
 for small and large $\Delta{}t/t_0$. 
The data for $t_0=0.1$ are omitted because of the strong finite-time effect. 
(e) Distribution of $\Delta{}q$ for given pairs of $t_0$ and $\Delta{}t/t_0$.
}
 \label{fig1}
\end{figure*}%

Such a stationary regime is, however, never attained within a finite time
 in an infinitely large system, unless a stationary interface is
 artificially given as an initial condition.
This is readily seen by recalling that the correlation length
 grows as $\xi\sim{}t^{2/3}$,
 whereas it is infinite for the stationary interfaces.
Therefore, in practical situations
 starting from a smooth or uncontrolled initial profile,
 one needs to elucidate how the interfaces approach the stationary regime
 in the course of time.
This is achieved by the present Letter.
Simulations of the PNG model
 show that the height difference of the flat interfaces
 exhibits a transition from the flat, growing regime to the stationary one.
We find scaling functions describing this crossover
 and determine their functional forms,
 which smoothly connect the GOE-TW and Baik-Rains $F_0$ distributions.
We also study the two-point correlation function
 and show how those for the two subclasses interplay at finite times.
These results are quantitatively reproduced by experimental data
 on turbulent liquid crystal
 \cite{Takeuchi.Sano-PRL2010,*Takeuchi.etal-SR2011,Takeuchi.Sano-JSP2012},
 indicating that they are universal characteristics
 of the KPZ-class interfaces.

First we study the PNG model.
Starting from a flat substrate $h(x,0)=0$,
 an interface experiences random nucleation events
 at a uniform rate.
On each nucleation, the local height $h(x,t)$ increases by one,
 producing a plateau that expands laterally at constant speed.
When two plateaux encounter, they simply coalesce.
For the simulations in continuous space and time,
 we numerically implement
 space-time representation used for analytical derivation
 of the distribution function
 \cite{Baik.Rains-JSP2000,Prahofer.Spohn-PRL2000},
 with the nucleation rate 2 per unit space and time
 and the plateau expansion speed 1.
This choice of the parameters corresponds to
 $v_\infty=2,A=2$, and $\Gamma=1$.
We impose the periodic boundary condition
 with system size $L=10^3$ and realize $10^4$ independent interfaces
 up to time $10^4$.
The size is chosen to satisfy $L\gg{}L_*$ until $t=10^4$ [\eqref{eq:FV}]
 so that the system does not reach the saturated regime,
 which is \textit{not} the crossover addressed in this Letter.

The quantity of interest is the height difference
 $\Delta{}h(x,\Delta{}t,t_0)\equiv{}h(x,t_0+\Delta{}t)-h(x,t_0)$,
 rescaled here as
\begin{equation}
 \Delta q(x,\Delta t, t_0) \equiv \frac{\Delta h - v_\infty \Delta t}{(\Gamma \Delta t)^{1/3}}.  \label{eq:QDef}
\end{equation}
By construction,
 $\Delta{}q\stackrel{\mathrm{d}}{\to}\chi_1$
 for $t_0\to0$ and then $\Delta{}t\to\infty$,
 while $\Delta{}q\stackrel{\mathrm{d}}{\to}\chi_0$
 for $t_0\to\infty$ and then $\Delta{}t\to\infty$,
 where $\chi_1$ and $\chi_0$ are random variables
 obeying the GOE-TW and Baik-Rains $F_0$ distributions, respectively,
 with the factor $2^{-2/3}$ multiplied with the usual definition
 for the former \cite{Baik.Rains-JSP2000,Prahofer.Spohn-PRL2000}.
Figure \ref{fig1}(a) shows the first- to fourth-order cumulants
 of $\Delta{}q$, $\cum{\Delta{}q^n}$,
 as functions of $\Delta{}t$ for different $t_0$,
 displayed with the values for the GOE-TW
 and Baik-Rains $F_0$ distributions (dashed and dotted lines, respectively).
The cumulants agree with those for the GOE-TW distribution
 as $\Delta{}t$ tends to infinity,
 while they indicate the values of the Baik-Rains $F_0$ distribution
 for large $t_0$ and small enough $\Delta{}t$.
The transition curves are found to collapse very well
 when $\Delta{}t$ is scaled by $t_0$ [\figref{fig1}(b)],
 except for
 too small $t_0$ and $\Delta{}t$.
In particular, for $t_0\to\infty$, the cumulants converge
 to a single set of functions,
 $\cum{\Delta{}q^n}\to\Delta{}Q_n(\Delta{}t/t_0)$,
 satisfying $\Delta{}Q_n(\tau)\to\cum{\chi_1^n}$
 for $\tau\to\infty$
 and $\Delta{}Q_n(\tau)\to\cum{\chi_0^n}$
 for $\tau\to0$.
One can indeed draw the functions $\Delta{}Q_n(\tau)$
 by making histograms for $\cum{\Delta{}q^n}$ at each $\Delta{}t/t_0$
 with varying $t_0$
 and fitting their modes
 by, e.g., spline functions,
 as shown by the black solid lines in \figref{fig1}(b).
Theoretical expressions of $\Delta{}Q_n(\tau)$ are unknown,
 because they involve time correlation
 which still remains analytically unsolved.
Asymptotically, the data suggest
 $\cum{\chi_0}-\Delta{}Q_1(\tau)\sim\tau^{2/3}$,
 $\cum{\chi_0^2}-\Delta{}Q_2(\tau)\sim\tau^{1/2}$ for small $\tau$ and
 $\Delta{}Q_1(\tau)-\cum{\chi_1}\sim\tau^{-1/3}$,
 $\Delta{}Q_2(\tau)-\cum{\chi_1^2}\sim\tau^{-2/3}$ for large $\tau$
 [\figref{fig1}(c,d)].
While this convergence to the GOE-TW distribution ($\tau\to\infty$)
 is analogous to that of the height variable $h(x,t)$
 \cite{Ferrari.Frings-JSP2011,Oliveira.etal-PRE2012,Takeuchi.Sano-JSP2012},
 the power laws toward the Baik-Rains $F_0$ distribution ($\tau\to0$)
 indicate unusual exponents that need to be explained theoretically.
For higher orders $n\geq3$, one needs
 better statistical accuracy to determine
 the asymptotics.
In between the two limits, the transition occurs earlier
 for larger $n$ ($\leq4$),
 leading to interesting undershoot
 in the skewness $\cum{\Delta{}q^3}/\cum{\Delta{}q^2}^{3/2}$
 and the kurtosis $\cum{\Delta{}q^4}/\cum{\Delta{}q^2}^2$
 [insets of \figref{fig1}(b)].
Finally, this crossover
 can also be checked directly in the distribution;
 \figref{fig1}(e) shows that the probability density functions of $\Delta{}q$
 overlap for fixed $\Delta{}t/t_0$,
 and that they shift from the Baik-Rains $F_0$ to the GOE-TW distributions
 as $\Delta{}t/t_0$ is increased.

\begin{figure}[t]
 \centering
 \includegraphics[width=\hsize,clip]{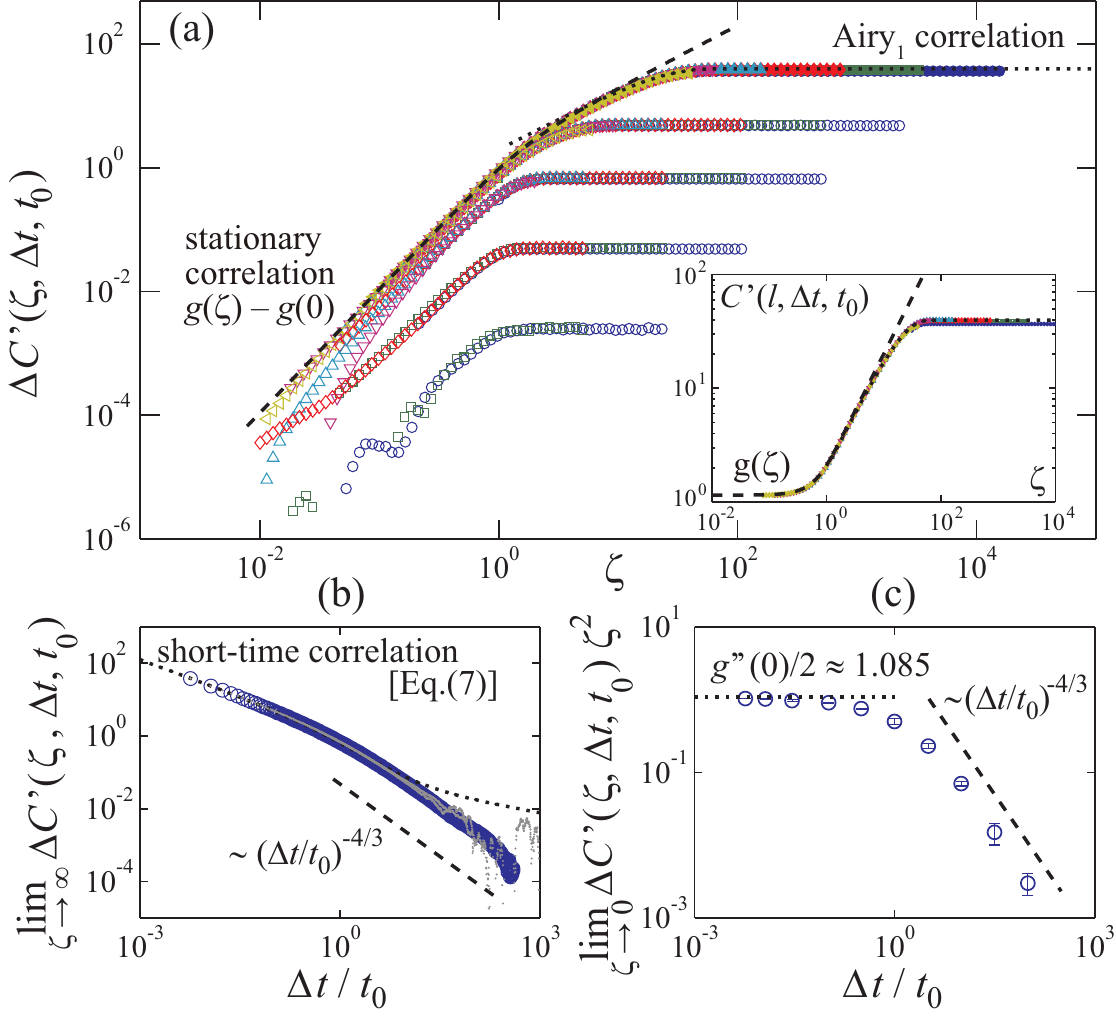}
 \caption{
(color online). 
Crossover in the correlation function for the PNG model. 
(a) $\Delta{}C'(\zeta,\Delta{}t,t_0)$
 against $\zeta$ for $\Delta{}t/t_0=0.006,0.1,1,10,100$
 (different sets of data; increasing from top to bottom)
 and $t_0=1,10,100,1000,4000,7000$
 (different colors and symbols; increasing from right to left). 
The dashed and dotted lines indicate
 the stationary correlation
 and the Airy$_1$ correlation
 as described in \eqref{eq:CorrFuncAsymp1},
 respectively, with $\Delta{}t/t_0=0.006$ for the latter.
The inset shows the data for $\Delta{}t/t_0=0.006$
 without subtraction of $C'(0,\Delta{}t,t_0)$.
(b,c) Asymptotics of $\Delta{}C'(\zeta,\Delta{}t,t_0)$
 for $\zeta\to\infty$ (b) and $\zeta\to0$ (c). 
Blue circles are the numerical estimates,
 obtained from the data with $t_0=10$ for (b)
 and with various $t_0$ for (c). 
The gray dots in (b) show the values of the right-hand side of \eqref{eq:Corr2ndCum},
 where $\cum{\Delta{}q^2}$ for $t_0=10$ and $100$ is used as $Q_2(\tau)$.
}
 \label{fig2}
\end{figure}

Now we turn our attention to the two-point correlation function,
 defined here by
\begin{equation}
 C(l, \Delta t, t_0) \equiv \Expct{[ \delta h(x+l, t_0 + \Delta t) - \delta h(x, t_0) ]^2} \label{eq:CorrFuncDef}
\end{equation}
 with $\delta{}h(x,t)\equiv{}h(x,t)-\expct{h(x,t)}$.
If one takes the stationary limit $t_0\to\infty$
 and then considers large $\Delta{}t$, one has
 $C'(\zeta,\Delta{}t,t_0)\equiv(\Gamma\Delta{}t)^{-2/3}C(l,\Delta{}t,t_0)\simeq g(\zeta)$
 with rescaled length $\zeta\equiv(Al/2)(\Gamma\Delta{}t)^{-2/3}$,
 where $g(\zeta)$ is the exact solution for the rescaled stationary correlation
 \cite{Prahofer.Spohn-JSP2004,Imamura.Sasamoto-PRL2012}.
This is tested in \figref{fig2}(a)
 with finite $t_0$ and $\Delta{}t$,
 where $\Delta{}C'(\zeta,\Delta{}t,t_0)\equiv{}C'(\zeta,\Delta{}t,t_0)-C'(0,\Delta{}t,t_0)$ is compared with $g(\zeta)-g(0)$ in the main panel.
First we note that the data for fixed $\Delta{}t/t_0$ and different $t_0$
 overlap with each other,
 confirming that $\Delta{}t/t_0$ is the only time scale
 that controls the dynamics.
Now, we focus on the data with the smallest $\Delta{}t/t_0$ we have,
 namely $\Delta{}t/t_0=0.006$,
 shown by solid symbols in \figref{fig2}(a) (top data set).
They are found to indicate the stationary correlation function
 for small $\zeta$,
 with or without subtraction of $C'(0,\Delta{}t,t_0)$
 (main panel and inset, respectively).
By contrast, for large $\zeta$, the correlation is governed
 by the spatial correlation of the flat interfaces,
 namely the Airy$_1$ correlation $g_1(\cdot)$,
 defined by $g_1(v)\equiv\expct{\mathcal{A}_1(u+v)\mathcal{A}_1(u)}-\expct{\mathcal{A}_1(u)}^2$
 with the Airy$_1$ process $\mathcal{A}_1(u)$
 \cite{Sasamoto-JPA2005,Borodin.etal-CMP2008,Note1}.
To see this, we take $\Delta{}t\to0$ in \eqref{eq:CorrFuncDef}
 and obtain, for large $t_0$,
$\Delta{}C'(\zeta,0,t_0)=C'(\zeta,0,t_0)\simeq2(\Delta{}t/t_0)^{-2/3}\[g_1(0)-g_1\((\Delta{}t/t_0)^{2/3}\zeta\)\]$.
This function with $\Delta{}t/t_0=0.006$ is indicated
 by the dotted line in \figref{fig2}(a)
 and accounts for the data with large $\zeta$.
In short, when $\Delta{}t/t_0$ is small enough,
\begin{equation}
 C'(\zeta, \Delta t, t_0) \!\simeq\! \begin{cases} g(\zeta) & \!\!(\zeta \ll \zeta_\mathrm{c}), \\ \! 2 (\tfrac{\Delta t}{t_0})^{-\tfrac{2}{3}} \! \[ g_1(0) \!-\! g_1 \! \( \!\! (\tfrac{\Delta t}{t_0})^{\tfrac{2}{3}} \zeta \! \)\] & \!\!(\zeta \gg \zeta_\mathrm{c}), \end{cases}  \label{eq:CorrFuncAsymp1}
\end{equation}
 where the crossover length $\zeta_\mathrm{c}$ is defined by the intersection
 of the two functions.
If $\Delta{}t/t_0$ is further decreased in \figref{fig2}(a),
 the Airy$_1$ branch
 moves away
 as $(\Delta{}t/t_0)^{-2/3}$ along both axes, leaving, asymptotically,
 only the stationary correlation $g(\zeta)$ as expected.
Alternatively,
 if $C$ and $l$ are rescaled by $t_0^{2/3}$ instead of $\Delta{}t^{2/3}$,
 what remains asymptotically is the Airy$_1$ correlation.
For tiny but finite $\Delta{}t/t_0$,
 the two branches are connected by $C'\simeq2\zeta$.

We then study how the correlation function varies for large $\Delta{}t/t_0$.
The data series in \figref{fig2}(a)
 show that $\Delta{}C'(\zeta,\Delta{}t,t_0)$ decreases
 with increasing $\Delta{}t/t_0$.
In the limit $\zeta\to\infty$,
 since $\expct{\delta{}h(x+l,t_0+\Delta{}t)\delta{}h(x,t_0)}\to0$,
 we have $\Delta{}C'(\zeta,\Delta{}t,t_0)\to2(\Gamma\Delta{}t)^{-2/3}C_\mathrm{t}(\Delta{}t,t_0)$
 with $C_\mathrm{t}(\Delta{}t,t_0)\equiv\expct{\delta{}h(x,t_0+\Delta{}t)\delta{}h(x,t_0)}$,
 i.e., the time correlation function.
Despite the lack of analytical solution,
 its short-time behavior ($\Delta{}t/t_0\ll1$) is given by
\begin{equation}
 C_\mathrm{t}(\Delta t, t_0) \simeq (\Gamma^2 t_0 t_\mathrm{r})^{\tfrac{1}{3}}\expct{\chi_1^2}_c \biggl[ 1 - \frac{\expct{\chi_0^2}_c}{2 \expct{\chi_1^2}_c} \!\( 1 - \frac{t_0}{t_\mathrm{r}} \)^{\!\!\tfrac{2}{3}} \biggr]  \label{eq:CorrShortTime}
\end{equation}
 with $t_\mathrm{r}\equiv{}t_0+\Delta{}t$
 \cite{Krug.etal-PRA1992,Kallabis.Krug-EL1999,Takeuchi.Sano-JSP2012}.
For $\Delta{}t/t_0\gg1$,
 numerical \cite{Kallabis.Krug-EL1999}
 and experimental \cite{Takeuchi.Sano-JSP2012} studies showed
 $C_\mathrm{t}(\Delta{}t,t_0)\simeq(\Gamma^2t_0t_\mathrm{r})^{1/3}F(\Delta{}t/t_0)$ with $F(\tau)\sim\tau^{-1}$.
They indicate
\begin{equation}
 \lim_{\zeta\to\infty}\! \Delta C'(\zeta, \Delta t, t_0) \sim \begin{cases} (\Delta t/t_0)^{-2/3}, & \!(\Delta t/t_0 \ll 1), \\ (\Delta t/t_0)^{-4/3}, & \!(\Delta t/t_0 \gg 1), \end{cases}  \label{eq:CorrShortLength}
\end{equation}
 and correctly account for the data [\figref{fig2}(b)].
Further, since the second-order cumulant of the rescaled height difference,
 $Q_2(\Delta{}t/t_0)$, involves the two-point time correlation
 $C_\mathrm{t}(\Delta{}t,t_0)$,
 we also obtain for arbitrary $\Delta{}t/t_0$
\begin{align}
 &\lim_{\zeta\to\infty} \Delta C'(\zeta, \Delta t, t_0) \notag \\
 &= \expct{\chi_1^2}_c \biggl[ \!\( 1+\frac{1}{\Delta t/t_0} \)^{\!\!\tfrac{2}{3}} \!\!+ (\Delta t/t_0)^{-\tfrac{2}{3}} \biggr] \!- Q_2(\Delta t/t_0).  \label{eq:Corr2ndCum}
\end{align}
This is also confirmed
 as shown by gray dots in \figref{fig2}(b).

In contrast to the long-length limit,
 one cannot \textit{a priori} predict
 how the short-length limit $\zeta\to0$
 of $\Delta{}C'(\zeta,\Delta{}t,t_0)$
 depends on $\Delta{}t/t_0$.
The data in \figref{fig2}(a) suggest
 $\Delta{}C'(\zeta,\Delta{}t,t_0)\sim\zeta^2$
 for any $\Delta{}t/t_0$.
Figure \ref{fig2}(c) shows that the coefficient of this quadratic term
 varies as
\begin{align}
 &\lim_{\zeta\to 0} \Delta C'(\zeta,\Delta t, t_0) \zeta^{-2} = \frac{1}{2}\left. \prts{C'}{\zeta}{2} \right|_{\zeta = 0} \notag \\
 &\qquad\qquad \simeq \begin{cases} g''(0)/2 \approx 1.085 & (\Delta t/t_0 \ll 1), \\ c (\Delta t/t_0)^{-4/3} & (\Delta t/t_0 \gg 1), \end{cases}    \label{eq:CorrLongLength}
\end{align}
 with a constant $c$ and the second derivative $g''(0)$,
 which naturally arises since $C'(\zeta,\Delta{}t,t_0)\to{}g(\zeta)$
 for $\Delta{}t/t_0\to0$.
To examine the other limit,
 let us note
 $\frac{1}{2}\left.\frac{\p^2C'}{\p\zeta^2}\right|_{\zeta=0}=(A/2)^{-2}(\Gamma\Delta{}t)^{2/3}\expct{\prt{h}{x}(x,t_0+\Delta{}t)\prt{h}{x}(x,t_0)}$,
 which is simply time correlation in the slope of the interface.
It is suggestive that
 the short- and long-length limits of $\Delta{}C'(\zeta,\Delta{}t,t_0)$
 are governed by the slope-slope and height-height time correlations,
 respectively, decaying with the same power
 in the rescaled units
 [\eqsref{eq:CorrShortLength} and \pref{eq:CorrLongLength}].
The results may also remind us of the space-like and time-like paths
 argued in the literature \cite{Ferrari-JSM2008,Corwin.etal-AIHPBPS2012},
 though precise relation
 is yet to be clarified.

Finally, we test universality of the presented crossover,
 analyzing experimental data of growing interfaces
 in turbulent liquid crystal.
While the readers are referred
 to Refs.~\cite{Takeuchi.Sano-PRL2010,Takeuchi.Sano-JSP2012}
 for detailed descriptions,
 in this series of work
 the author and a coworker studied planar evolution of borders
 between two distinct regimes of spatiotemporal chaos,
 called the dynamic scattering modes 1 and 2,
 in the electroconvection of nematic liquid crystal.
The interfaces grow under high applied voltage,
 clearly exhibiting, besides the exponents,
 the distribution and correlation functions
 for the flat and curved KPZ-class interfaces
 \cite{Takeuchi.Sano-PRL2010,Takeuchi.Sano-JSP2012}.
Here, we employ the data for 1128
 flat interfaces used in Ref.~\cite{Takeuchi.Sano-JSP2012}
 and perform the crossover analyses developed in the present study.

\begin{figure}[t]
 \centering
 \includegraphics[width=\hsize,clip]{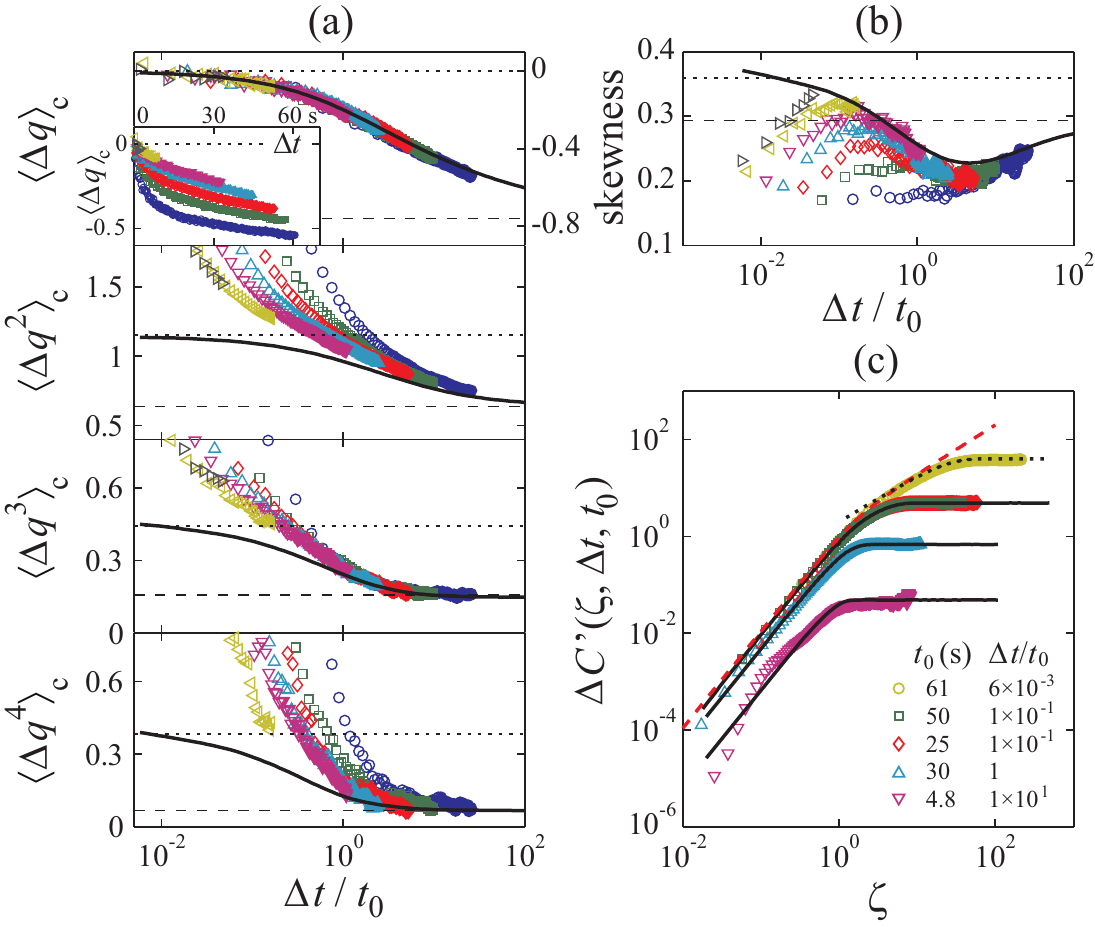}
 \caption{
(color online). 
Crossover in the liquid-crystal experiment. 
(a,b) Cumulants (a)
 and skewness (b)
 against $\Delta{}t/t_0$ with $t_0=2,6,10,18,30,54,60\unit{s}$
 (from right to left). 
The inset shows $\cum{\Delta{}q}$ against $\Delta{}t$. 
The dotted and dashed lines indicate
 the values for the Baik-Rains $F_0$ and GOE-TW distributions, respectively. 
The solid curves show the fitting to the PNG data
 obtained in \figref{fig1}(b). 
(c) Rescaled correlation function $\Delta{}C'(\zeta,\Delta{}t,t_0)$
 against $\zeta$ for given $t_0$ and $\Delta{}t/t_0$
 ($\Delta{}t/t_0$ increases from top to bottom). 
The red dashed and black dotted lines indicate
 the stationary and Airy$_1$ correlation functions
 as described in \eqref{eq:CorrFuncAsymp1}, respectively,
 the latter being set with $\Delta{}t/t_0=0.006$. 
The three black solid lines trace PNG data in \figref{fig2}(a)
 with corresponding values of $\Delta{}t/t_0$.
}
 \label{fig3}
\end{figure}%

Figure \ref{fig3} shows the results.
The $n$th-order cumulants of the rescaled height difference $\Delta{}q$
 [\eqref{eq:QDef}] with various $t_0$,
 which sufficiently fall apart as functions of $\Delta{}t$
 [see, e.g., inset of \figref{fig3}(a)],
 collapse reasonably well when plotted against $\Delta{}t/t_0$
 [\figref{fig3}(a)],
 despite a rather strong finite-time effect for $n\geq2$.
The collapsed data are found asymptotically
 on top of the fitting curves obtained for the PNG model,
 $Q_n(\Delta{}t/t_0)$ (black solid lines).
This implies that $Q_n(\tau)$ are universal functions
 of the KPZ class describing
 the crossover in question,
 and so is the distribution function of $\Delta{}q$
 parametrized by $\Delta{}t/t_0$.
The undershoot in the skewness is also confirmed experimentally
 [\figref{fig3}(b)], while it was not clearly identified
 for the kurtosis because of larger statistical error (not shown).
Moreover, extrapolation of the finite-time corrections
 in the cumulants allows us to roughly estimate
 the time needed for direct observation of the Baik-Rains $F_0$ distribution,
 longer than $10^3\unit{s}$ here,
 which is unfortunately unreachable in the current setup
 \cite{Takeuchi.Sano-PRL2010,Takeuchi.Sano-JSP2012}.

The results on the correlation function are also reproduced experimentally
 [\figref{fig3}(c)].
The functional form is parametrized solely by $\Delta{}t/t_0$
 (see two data sets for $\Delta{}t/t_0=10^{-1}$
 overlapping with each other)
 and agrees very well with the one obtained for the PNG model
 (black solid lines).
In particular, the crossover
 between the stationary and Airy$_1$ correlations
 [\eqref{eq:CorrFuncAsymp1}] is clearly confirmed
 for small enough $\Delta{}t/t_0$ (top yellow data set).

In summary, we have studied
 the flat-stationary crossover in the KPZ class,
 which takes place gradually in time.
Analyzing numerical and experimental data,
 we have found and determined
 universal functions describing
 the cumulants and the two-point correlation during this crossover.
These functions show
 multifaceted relations to the analytically unsolved time correlation,
 and hence may provide an important clue toward its solution.
Seeking a possible connection to analogous,
 mathematically tractable crossover in space \cite{Corwin-RMTA2012}
 is another interesting issue left for future studies.
Besides such fundamental importance,
 our results also answer a practical question of
 how interfaces realized in experiments and simulations
 approach the stationary regime, 
 which is never attained without full control on the initial condition.

\begin{acknowledgments}
The author acknowledges
 enlightening suggestions by T. Sasamoto and H. Spohn
 during the MSRI workshop in 2010
 ``Random Matrix Theory and its Applications II,''
 which gave birth to the present work.
Fruitful discussions with them and T. Imamura are also appreciated,
 as well as a remark by Y. Nakayama on numerical implementation
 of the PNG model.
Further, the author thanks M. Pr\"{a}hofer
 for providing the theoretical curve
 of the GOE-TW distribution,
 T. Imamura for those of the Baik-Rains $F_0$ distribution
 and the stationary correlation function $g(\zeta)$,
 and F. Bornemann for that of the Airy$_1$ correlation function $g_1(\zeta)$
 \cite{Bornemann-MC2010}.
This work is supported in part
 by Grant for Basic Science Research Projects from The Sumitomo Foundation.
\end{acknowledgments}

\bibliography{KPZstat}

\end{document}